\documentclass[letterpaper, 11pt]{article}
\usepackage[T1]{fontenc}
\usepackage{lmodern}
\usepackage{listings}

\usepackage{graphicx}
\usepackage{authblk}
\usepackage{amsmath}
\usepackage{geometry}
\usepackage{parskip}
\usepackage{setspace}
\usepackage{caption}
\usepackage{xcolor}
\usepackage{hyperref}
\usepackage{booktabs}
\usepackage{comment}
\usepackage{afterpage}
\usepackage{chemformula}
\usepackage{adjustbox}
\usepackage{pifont}
\usepackage{multirow}
\usepackage{orcidlink}
\usepackage{algorithm}      
\usepackage{algpseudocode}  
\usepackage{float} 
\usepackage{placeins}
\newcommand{\fplus}{\mathord{\text{\ding{58}}}}

\hypersetup{
    colorlinks=true,
    linkcolor=black,   
    urlcolor=teal,
    citecolor=blue
}

\title{\textbf{pyzentropy: A Python package implementing recursive entropy for first-principles thermodynamics}}
\author{\small Nigel Lee En Hew\orcidlink{0000-0003-1374-4589}\textsuperscript{1,*}, 
Luke Allen Myers\orcidlink{0009-0003-0823-0871}\textsuperscript{1}, 
Shun-Li Shang\orcidlink{0000-0002-6524-8897}\textsuperscript{1}, 
Zi-Kui Liu\orcidlink{0000-0003-3346-3696}\textsuperscript{1}}
\affil{\small \textsuperscript{1}Department of Materials Science and Engineering, The Pennsylvania State University, University Park, Pennsylvania 16802, USA}
\affil{\small \textsuperscript{*}Corresponding author: njh5724@psu.edu}

\date{} 

\begin{document}

\maketitle

\vspace{-2em}  

\section*{Abstract}
While the recursive property of entropy is well known in information theory, it is rarely utilized in thermodynamics, despite entropy originating in this field. Moreover, computational tools to implement this concept within first-principles thermodynamics remain lacking. In this work, we introduce an open-source Python package, \texttt{pyzentropy}, to implement this approach. We demonstrate its effectiveness using \ch{Fe3Pt} as a case study, considering a 12-atom supercell with multiple magnetic configurations. By applying the recursive formulation of entropy to compute the total entropy of the system, we reproduce the Invar behavior, along with the anomalous temperature dependence of the linear coefficient of thermal expansion (LCTE), heat capacity $C_P$, and bulk modulus $B$. We also construct the $T$–$V$ and $P$–$T$ phase diagrams in good agreement with experimental observations. Finally, we highlight the importance of determining key high-probability configurations to accurately capture material properties.

\section{Introduction}
Entropy is a macroscopic thermodynamic quantity that is defined by the probability distribution over a system’s microstates, quantifying the system’s microscopic uncertainty. From entropy, other macroscopic thermodynamic properties can be derived. The well-known expression introduced by Boltzmann, famously engraved on his tombstone, is commonly written as 
\begin{equation}\label{eq:boltzmann_entropy}
    S = k_B \ln{\Omega},
\end{equation}
where $k_B$ is Boltzmann's constant, and $\Omega$ is the number of accessible microstates corresponding to the particular macrostate of the system. Eq.~\eqref{eq:boltzmann_entropy} assumes that all accessible microstates are equally probable. Gibbs later generalized this expression to account for non-equal probabilities: 
\begin{equation} \label{eq:gibbs_entropy}
    S = -k_B\sum_{x}  p_x \ln p_x, 
\end{equation}
where $p_x$ is the probability of microstate $x$. If all probabilities are equal, Eq.~\eqref{eq:gibbs_entropy} reduces to Eq.~\eqref{eq:boltzmann_entropy}. Eq.~\eqref{eq:gibbs_entropy} is the definition of the Gibbs entropy used in statistical mechanics, and the same mathematical form without $k_B$ is known as the Shannon entropy in information theory. 

For a single atomic configuration $k$, such as that used in density functional theory (DFT) calculations, there exist well-known entropies that can be computed. Using Bose-Einstein statistics, the vibrational entropy of $k$, $S_{k,vib}$, can be expressed as
\begin{align}
S_{k,vib} &= -k_B \sum_{\mathbf{q}\nu} 
\Big[n_{\mathbf{q}\nu} \ln n_{\mathbf{q}\nu}
- (1+n_{\mathbf{q}\nu}) \ln(1+n_{\mathbf{q}\nu})
\Big] \label{eq:Svib_BE} \\
&= \frac{1}{2T} 
\sum_{\mathbf{q}\nu} 
\hbar\omega_{\mathbf{q}\nu}
\coth\left( \frac{\hbar\omega_{\mathbf{q}\nu}}{2k_B T} \right)
- k_B
\sum_{\mathbf{q}\nu}
\ln\left[ 2\sinh\left( \frac{\hbar\omega_{\mathbf{q}\nu}}{2k_B T} \right) \right]. \label{eq:Svib_qv}
\end{align}
Eq.~\eqref{eq:Svib_BE} expresses the vibrational entropy in terms of the mean occupation number $n_{q\nu}$ of phonon modes $\{\mathbf{q},\nu\}$, as described by the Bose-Einstein distribution function. It is also commonly expressed in the form of Eq.~\eqref{eq:Svib_qv}, where $\omega_{q\nu}$ is the phonon frequency of the phonon mode $\{\mathbf{q},\nu\}$, and $\hbar$ is the reduced Planck constant. Simpler approximations, such as the Debye--Grüneisen model, may also be employed when a full phonon calculation is computationally prohibitive.

In a similar manner, using Fermi-Dirac statistics, the electronic entropy $S_{k, el}$ can be expressed as
\begin{align}
S_{k,el} &= -k_B \sum_i \Big[ n_i \ln n_i + (1-n_i) \ln(1-n_i) \Big], \label{eq:Sel_i}
\end{align}
 where $n_i$ is the mean occupation number of an electronic state $i$ described by the Fermi-Dirac distribution function. More details regarding Eqs.~\eqref{eq:Svib_BE}–\eqref{eq:Sel_i} are well-documented and can be found in standard textbooks (e.g., \cite{Dove1993IntroductionDynamics, Grosso2014SolidPhysics, beale2011statistical}).

It is commonly assumed that the vibrational and electronic degrees of freedom are statistically independent. Under this assumption, the total entropy of configuration $k$ can be written as the sum of the two contributions: 
\begin{align} \label{eq:gibbs_entropy_all_vib_el}
S_k &= S_{k,vib} + S_{k,el}.
\end{align}
We also refer to $S_k$ as the \emph{intra-configurational entropy}, i.e., the entropy associated with a single atomic configuration $k$.

Now, consider a system which comprises $N$ distinct atomic configurations. To calculate the total entropy of the system, we employ the recursion property of entropy from information theory \cite{Bengtsson2017GeometryDistributions}, which involves coarse-graining or grouping microstates $x$ into configurations $k$. Doing so allows us to rewrite Eq.~\eqref{eq:gibbs_entropy} into the equation below:
\begin{align} \label{eq:zentropy}
    S &= -k_B \sum_{k=1}^{N} p_k \ln p_k + \sum_{k=1}^{N} p_k S_k,
\end{align}
where $k$ indexes 1 to $N$ configurations and $p_k$ is the probability of configuration $k$. We refer to Eq.~\eqref{eq:zentropy} as the entropy equation in the \emph{zentropy} approach. The term zentropy combines the German \emph{Zustandssumme} (“sum over states”) with entropy, emphasizing that it involves summing over states or configurations, each possessing its own intrinsic entropy.

In Eq.~\eqref{eq:zentropy}, the first term is commonly referred to as the \emph{configurational entropy}, while the second term is the probability-weighted sum of the intra-configurational entropies. Neglecting the second term implicitly treats each configuration $k$ as a single microstate, thereby retaining only the configurational entropy. However, each configuration $k$ represents a \emph{group of microstates}, which introduces an additional entropy contribution captured by the second term.

While the recursion property of entropy is well-known in information theory, the concept is relatively unknown in thermodynamics, despite thermodynamics being the field where entropy originated. This work highlights the importance of considering different perspectives from different fields. Analogous to the parable of the blind men and the elephant, information theorists and thermodynamicists approach entropy from different viewpoints and would benefit from integrating their insights \cite{Perdew2021InterpretationsTheories}.

The goal of this work is to develop \texttt{pyzentropy}, an open-source tool for efficiently implementing this approach. In the following section, we review the methodology underlying this approach, including the formalism and its implementation in \texttt{pyzentropy}. Using configuration properties as inputs, the package computes $(V, T)$- and $(P, T)$-dependent properties for the system, enabling the construction of $V$-$T$ and $P$-$T$ phase diagrams. Finally, we demonstrate the application of \texttt{pyzentropy} to the \ch{Fe3Pt} system. We show the package’s ability to compute thermodynamic properties, including the prediction of negative thermal expansion, and to generate publication-ready figures. While \ch{Fe3Pt} has been studied previously \cite{Wang2010ThermodynamicPrototype, Shang2023QuantifyingFe3Pt}, we use it here as a case study to demonstrate the capabilities of this new package, providing new plots and additional insights.

\section{Methodology}
\subsection{Formalism}

We use Eq.~\eqref{eq:zentropy} as a starting point and apply the principle of maximum entropy~\cite{Myers2025RecursiveApproach} under the constraints that the configuration probabilities $p_k$ sum to one,
\begin{equation} \label{eq:probability_constraint}
    \sum_{k=1}^{N} p_k = 1,
\end{equation}
and that the internal energy is fixed,
\begin{equation}
    \langle E \rangle = \sum_{k=1}^{N} p_k E_k.
\end{equation}
Here, $E_k$ is the internal energy of configuration $k$, and $\langle E \rangle$ is the ensemble-averaged internal energy. The resulting partition function $Z$ and configuration probability $p_k$ are then
\begin{equation} \label{eq:partition_function}
    Z = \sum_{k=1}^{N} Z_k = \sum_{k=1}^{N} \exp\!\left(-\frac{F_k}{k_B T}\right),
\end{equation}
\begin{equation} \label{eq:probability_k}
    p_k = \frac{\exp\!\left(-\frac{F_k}{k_B T}\right)}{Z},
\end{equation}
where $Z_k$ is the partition function of configuration $k$, $F_k$ is its Helmholtz energy, and $T$ is the temperature. If Eq.~\eqref{eq:gibbs_entropy} were used as the starting point instead, i.e., without grouping microstates into configurations, the exponentials in Eqs.~\eqref{eq:partition_function} and~\eqref{eq:probability_k} would be functions of the typical microstate energies $E_x$ rather than the configuration Helmholtz energies $F_k$.

The Helmholtz energy of the system is given by
\begin{equation}\label{eq:fundamental_F}
    F = -k_B T \ln Z,
\end{equation}
which, upon algebraic manipulation, can also be expressed as
\begin{equation}\label{eq:derived_F}
    F = \sum_{k=1}^{N} p_k F_k + k_B T \sum_{k=1}^{N} p_k \ln p_k.
\end{equation}

The first and second derivatives of $F$ with respect to the volume $V$ at fixed temperature $T$ are
\begin{equation} \label{eq:dF/dV}
    \left(\frac{\partial F}{\partial V}\right)_T 
    = \sum_{k=1}^{N} p_k \left(\frac{\partial F_k}{\partial V}\right)_T,
\end{equation}
\begin{equation} \label{eq:d^2F/dV^2}
    \left(\frac{\partial^2 F}{\partial V^2}\right)_T 
    = \sum_{k=1}^{N} p_k \left(\frac{\partial^2 F_k}{\partial V^2}\right)_T
      + \frac{1}{k_B T} 
        \left[
            \left( \sum_{k=1}^{N} p_k \left(\frac{\partial F_k}{\partial V}\right)_T \right)^2
            - \sum_{k=1}^{N} p_k \left(\frac{\partial F_k}{\partial V}\right)_T^{\!2}
        \right].
\end{equation}

The heat capacity at constant volume $C_V$, and the bulk modulus $B$ of the system are
\begin{equation} \label{eq:heat_capacity_system}
    C_V 
    = \left(\frac{\partial \langle E \rangle}{\partial T}\right)_V
    = -T \left(\frac{\partial^2 F}{\partial T^2}\right)_V
    = \sum_{k=1}^{N} p_k C_{V,k} 
      + \frac{1}{k_B T^2} 
        \left[
            \sum_{k=1}^{N} p_k E_k^2 
            - \left(\sum_{k=1}^{N} p_k E_k \right)^2
        \right],
\end{equation}
\begin{equation} \label{eq:bulk_modulus_system}
    B = V \left(\frac{\partial^2 F}{\partial V^2}\right)_T.
\end{equation}

These quantities are functions of $V$ and $T$. Obtaining $F(V,T)$ in Eqs.~\eqref{eq:fundamental_F} and \eqref{eq:derived_F} enable the determination of thermal properties at constant pressure $P$. This is achieved by minimizing $F(V,T) + PV$ with respect to $V$ at fixed $T$ to obtain the Gibbs energy $G(T, P)$. Consequently, other system properties at constant $P$, such as the equilibrium volume $V_{eq}$ (the volume that minimizes the Gibbs energy), the coefficient of thermal expansion (CTE), $p_k$, $S$, $C_p$, and $B$, as well as $P$–$T$ and $T$–$V$ diagrams, can be obtained.

\subsection{Implementation in pyzentropy}
The pseudocode representing \texttt{pyzentropy}'s workflow is shown in Algorithm~\ref{alg:pseudocode}. It illustrates the procedure for computing thermodynamic properties from first-principles inputs. The workflow is organized around two main classes: \texttt{Configuration} and \texttt{System}. 

Each \texttt{Configuration} represents a distinct atomic configuration of the system and requires inputs for the Helmholtz energy $F_k$, its derivatives with respect to volume ($dF_k/dV$, $d^2F_k/dV^2$), as well as the entropy $S_k$ and heat capacity $C_{V,k}$ at a set of $(V, T)$ points. 
In practice, these inputs are attainable via the standard quasi-harmonic approximation, as implemented in several open-source packages \cite{Togo2023ImplementationPhono3py, Togo2023First-principlesPhono3py, Hew2025DensityApproximation}, where $F_k$ vs. $V$ at each fixed $T$ is fitted to an equation of state. 
From these inputs, the internal energy $E_k = F_k + T S_k$ for each \texttt{Configuration} is then calculated.

The \texttt{System} class aggregates all \texttt{Configuration}s to compute system properties using Eq.~(\ref{eq:zentropy}) and Eqs.~(\ref{eq:partition_function})--(\ref{eq:bulk_modulus_system}). 
These properties include the total partition function $Z$, configuration probabilities $p_k$, Helmholtz energy $F$, along with its first and second derivatives, bulk modulus $B$, entropy $S$, and heat capacity $C_V$ at $(V, T)$.

From then on, two primary calculation modes are implemented:
\begin{enumerate}
    \item \textbf{Calculate properties at fixed $P$:} This mode computes properties for a selected value of $P$. 
    For each $T$, the equilibrium volume $V_{eq}$ is obtained by minimizing $F + PV$, i.e., by solving $dF/dV + P = 0$, which yields the Gibbs energy $G$ at $(T, P)$. 
    Thermodynamic properties such as $S$, $B$, and $p_k$ are evaluated at $V_{eq}$ to obtain their respective values at $P$, 
    while CTE and the heat capacity at constant pressure $C_P$ are obtained from temperature derivatives. 
    This calculation mode provides these properties at a fixed $P$ for varying $T$.

    \item \textbf{Calculate phase diagrams:} This mode builds on the previous mode for calculating properties at different fixed $P$. 
    For each $P$, a second-order phase transition is identified when the ground-state probability $p_{GS} = 0.5$ (or when the sum of all other non-ground-state probabilities equals 0.5). 
    If this crossover exists, the corresponding $T$ and $V_{\rm eq}$ are recorded. 
    The resulting $(P, T, V_{eq})$ points are then taken as second-order transition points.
    
    The first-order transition points are determined using $F(V, T)$. 
    For each temperature $T$, a common tangent construction is attempted: two points on the $F$ vs. $V$ curve are identified such that the slope of the tangent line, given by $dF/dV$, is equal at both points and also equal to the slope of the secant line connecting them. 
    The pressure at the transition is then given by $P = -dF/dV$. 
    The resulting $(P, T, V)$ points are recorded as first-order transition points.
\end{enumerate}

For convenience, both the \texttt{Configuration} and \texttt{System} classes provide built-in plotting methods for visualizing all computed thermodynamic properties and phase-diagram data.

\subsection{Using pyzentropy} 
\texttt{pyzentropy} is an open-source Python package hosted on GitHub \cite{Pyzentropy} under the MIT license and is available through PyPI. The MIT license is a permissive open-source license that allows anyone to use, copy, modify, merge, publish, distribute, sublicense, and/or sell copies of the software, provided that the copyright and permission notice are included in all copies or substantial portions of the software. 

The package follows the semantic versioning scheme of MAJOR.MINOR.PATCH \cite{TomPreston-WernerSemantic2.0.0}. It includes automated tests to ensure the software works correctly and to help prevent future changes or additions from introducing errors, thereby maintaining long-term reliability. Furthermore, it is documented on Read the Docs \cite{NigelLeeEnHew2026PyzentropyDocumentation}, which provides detailed information about the API, installation instructions, and usage examples, including the example demonstrated in the following section.

\section{\ch{Fe3Pt} System}
Fe-Pt alloys are classic Invar materials, exhibiting very low or even negative thermal expansion coefficients below their Curie temperatures \cite{Sumiyama1979CharacteristicAlloys}. As a case study, we consider a 12-atom supercell of ordered \ch{Fe3Pt}, enumerate the symmetry-inequivalent magnetic configurations compatible with this supercell, and compute their properties, which are then used as inputs to determine the properties of the system.

\begin{algorithm}[H]
\caption{Pseudocode representation of the \texttt{pyzentropy} workflow using the \texttt{Configuration} and \texttt{System} classes. 
The term $p_{GS}$ denotes the probability of the ground-state configuration.}
\label{alg:pseudocode}

\begin{algorithmic}[1]

\For{each \texttt{Configuration}}
    \State Input $F_k$, $dF_k/dV$, $d^2F_k/dV^2$, $S_k$, and $C_{V,k}$ at $(V,T)$
    \State Calculate $E_k$ at $(V,T)$
\EndFor
\Statex\vspace{-0.5em} 

\For{the \texttt{System}}
    \State Input \texttt{Configuration}s
    \State Calculate $Z$, $p_k$, $F$, $dF/dV$, $d^2F/dV^2$, $B$, $S$, $C_v$ at $(V,T)$
    \Statex\vspace{-0.5em} 
    
    \Statex\hspace{\algorithmicindent}\textbf{Calculation options:}
    \Statex\hspace{\algorithmicindent}\textbf{(1) Calculate properties at fixed $P$}
    \State\hspace{\algorithmicindent}\textbf{for} each $T$ \textbf{do}
    \State\hspace{\algorithmicindent}\hspace{\algorithmicindent}Find global minimum of $F + PV$ (solve $dF/dV + P = 0$) to get $V_{\text{eq}}$ and $G$ at $(T,P)$
    \State\hspace{\algorithmicindent}\hspace{\algorithmicindent}Evaluate $S$, $B$, and $p_k$ at $V_{\text{eq}}$ to obtain properties at $(T,P)$
    \State\hspace{\algorithmicindent}\textbf{end for}
    \State\hspace{\algorithmicindent}Calculate CTE and $C_P$ at $(T,P)$:
    \State\hspace{\algorithmicindent}\hspace{\algorithmicindent}$CTE = \frac{1}{V_{\text{eq}}} \left(\frac{\partial V_{\text{eq}}}{\partial T}\right)_P$
    \State\hspace{\algorithmicindent}\hspace{\algorithmicindent}$C_P = T \left(\frac{\partial S}{\partial T}\right)_P$
    \Statex\vspace{-0.5em} 
    
    \Statex\hspace{\algorithmicindent}\textbf{(2) Construct phase diagrams}
    \Statex\hspace{\algorithmicindent}\textit{Second-order phase transition (ground-state crossover)}
    \State\hspace{\algorithmicindent}Initialize $P = 0$
    \State\hspace{\algorithmicindent}\textbf{while} True \textbf{do}
    \State\hspace{\algorithmicindent}\hspace{\algorithmicindent}Perform calculation (1) at $P$ 
    \State\hspace{\algorithmicindent}\hspace{\algorithmicindent}Find $T$ where $p_{GS}(T,P) = 0.5$
    \State\hspace{\algorithmicindent}\hspace{\algorithmicindent}\textbf{if} such $T$ exists \textbf{then}
        \State\hspace{\algorithmicindent}\hspace{\algorithmicindent}\hspace{\algorithmicindent}Find corresponding $V_{eq}$ at $(T,P)$
        \State\hspace{\algorithmicindent}\hspace{\algorithmicindent}\hspace{\algorithmicindent}Record $(P,T,V_{eq})$ as a second-order transition point
        \State\hspace{\algorithmicindent}\hspace{\algorithmicindent}\hspace{\algorithmicindent}Increment $P$ by $\Delta P$
    \State\hspace{\algorithmicindent}\hspace{\algorithmicindent}\textbf{else}
        \State\hspace{\algorithmicindent}\hspace{\algorithmicindent}\hspace{\algorithmicindent}Break loop
    \State\hspace{\algorithmicindent}\textbf{end while}
    \Statex\vspace{-0.5em} 
    
    \Statex\hspace{\algorithmicindent}\textit{First-order phase transition (miscibility gap)}
    \State\hspace{\algorithmicindent}\textbf{for} each $T$ \textbf{do}
    \State\hspace{\algorithmicindent}\hspace{\algorithmicindent}Attempt common tangent construction to find $V_1, V_2$ (equal $dF/dV$ and secant slope)
    \State\hspace{\algorithmicindent}\hspace{\algorithmicindent}\textbf{if} a valid tangent is found \textbf{then}
    \State\hspace{\algorithmicindent}\hspace{\algorithmicindent}\hspace{\algorithmicindent}Record $(P=-dF/dV, T, V_1, V_2)$ as a first-order transition point
    \State\hspace{\algorithmicindent}\hspace{\algorithmicindent}\textbf{else}
    \State\hspace{\algorithmicindent}\hspace{\algorithmicindent}\hspace{\algorithmicindent}Break loop
    \State\hspace{\algorithmicindent}\hspace{\algorithmicindent}\textbf{end if}
    \State\hspace{\algorithmicindent}\textbf{end for}

\EndFor
\end{algorithmic}
\end{algorithm}

\subsection{Generation of Input Data}
Density functional theory (DFT) calculations were performed using the Vienna \emph{ab initio} Simulation Package (VASP) \cite{Kresse1996EfficiencySet, Kresse1996EfficientSet} with the projector augmented-wave (PAW) method \cite{Kresse1999FromMethod, Blochl1994ProjectorMethod}. The Density Functional Theory ToolKit (\texttt{DFTTK})~\cite{Hew2025DensityApproximation} was used to enumerate all 37 symmetry-inequivalent collinear magnetic spin configurations of a 12-atom $1\times1\times3$ \ch{Fe3Pt} supercell (out of 512 possible configurations), considering only magnetic moments on Fe atoms. \texttt{DFTTK} was subsequently used to automate VASP calculations and perform post-processing to generate the input data required for \texttt{pyzentropy}.

\texttt{DFTTK} was used to compute $F_k$ for each configuration by summing three components:
\begin{equation}
F_k(V,T) = E_{k,0}(V) + F_{k,\mathrm{vib}}(V,T) + F_{k,\mathrm{el}}(V,T),
\label{eq:Fk_components}
\end{equation}
where $E_{k,0}(V)$ represents the static total energy at 0 K,
$F_{k,vib}(V, T)$ is the vibrational contribution to the Helmholtz energy computed using the Debye--Gr\"uneisen model with parameters $s = 0.617$ and $x = 2/3$, and $F_{k, el}(V, T)$ is the thermal electronic contribution to the Helmholtz energy \cite{Shang2010First-principlesNi3Al}. Of the 37 configurations, 25 were retained for subsequent analysis, as the remaining 12 relaxed into configurations already represented in the retained set.

The Perdew–Burke–Ernzerhof (PBE) exchange-correlation functional \cite{Perdew1996GeneralizedSimple} was used. PAW pseudopotentials were employed with the PBE.54 POTCAR files: Fe with valence configuration $3p^6 4s^2 3d^6$ and Pt with $5d^9 6s^1$. A gamma-centered $13 \times 13 \times 4$ k-point grid was used, corresponding to a density of approximately 8800 k-points per number of atoms. Key INCAR tags used in the calculations are summarized in \textbf{Table~\ref{tab:incar}}. More details on generating the input data can be found in Refs. \cite{Hew2025DensityApproximation, DFTTK}.

\begin{table}[htbp]
\centering
\caption{Key VASP INCAR tags used in the DFT calculations. The cell shape and atomic positions were relaxed at fixed volume in two consecutive runs, followed by a final static calculation. Relaxation and static settings are indicated where different.}
\label{tab:incar}
\begin{tabular}{lll}
\hline
\textbf{Tag} & \textbf{Value} & \textbf{Description} \\
\hline
EDIFF    & 1e-6                     & Electronic energy convergence criterion (eV) \\
EDIFFG   & -0.01                    & Ionic force convergence criterion (eV/Å) \\
ENCUT    & 520                      & Plane-wave cutoff energy (eV) \\
GGA      & PE                       & Exchange-correlation functional \\
IBRION   & 2 (relax); -1 (final)   & Ionic relaxation algorithm \\
ISIF     & 4                        & Stress and ionic relaxation type \\
ISMEAR   & 1 (relax); -5 (final)   & Smearing method \\
ISPIN    & 2                        & Spin polarization \\
LORBIT   & 11                       & DOS and projections \\
MAGMOM   & Set automatically by \texttt{DFTTK} & Initial magnetic moments (µB) \\
PREC     & Accurate                 & Precision setting \\
SIGMA    & 0.2 eV                   & Smearing width \\
\hline
\end{tabular}
\end{table}

\begin{figure}[!htbp]
    \centering
    \includegraphics[width=1\textwidth, trim=0.1cm 0.3cm 0cm 0.1cm, clip]{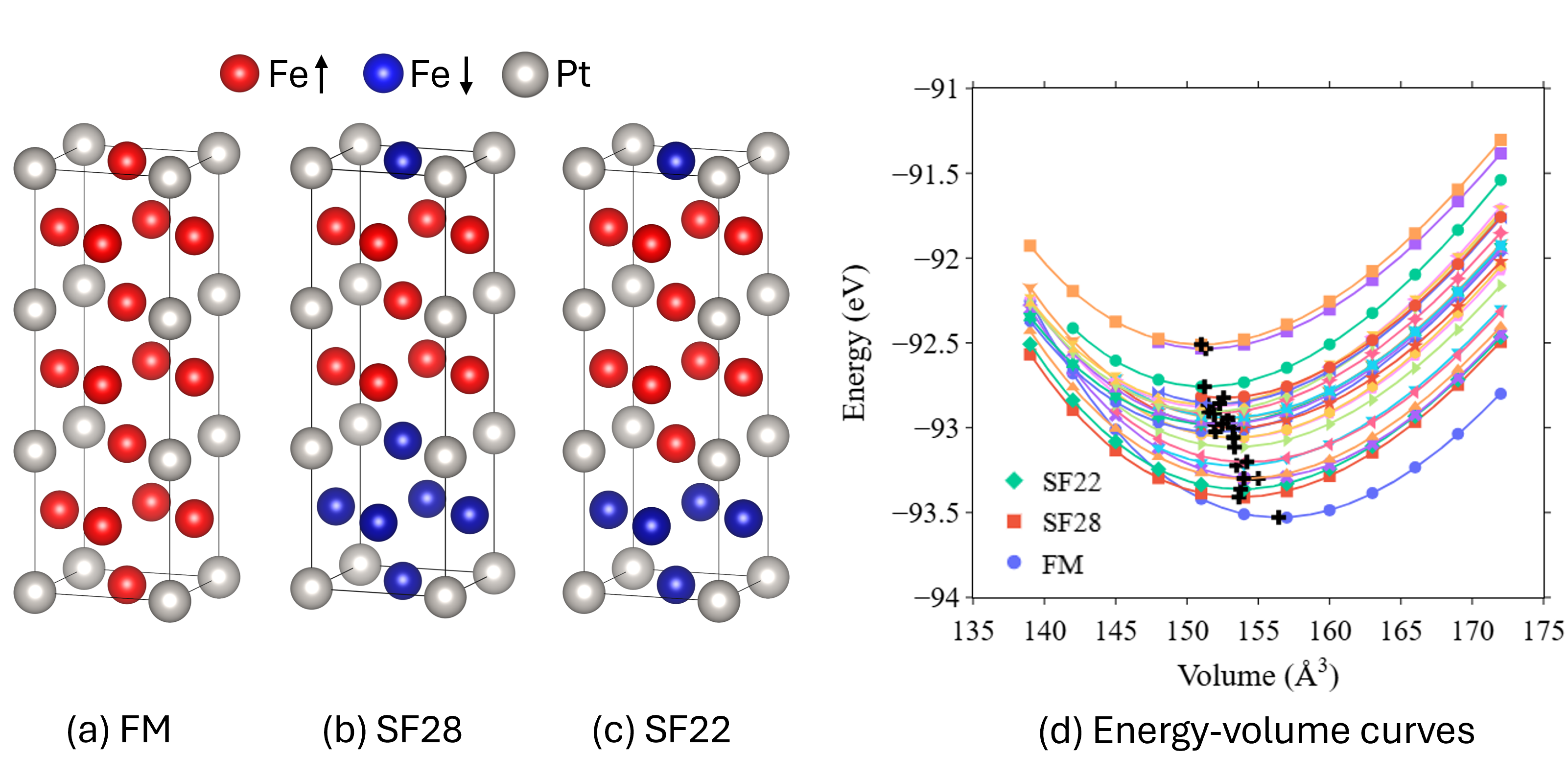}
    \caption{The three lowest energy $\text{Fe}_3\text{Pt}$ configurations using 12-atom supercells: (a) Ferromagnetic (FM) ground state, (b) spin-flipping (SF) 28, and (c) SF22. (d) Energy-volume curves for 25 $\text{Fe}_3\text{Pt}$ magnetic configurations using 12-atom supercells. The solid points represent the DFT data, while the curves are from the third-order or four-parameter Birch–Murnaghan equation of state fitted to those points. The $\fplus$ symbols denote the energy–volume minima. Only the three lowest energy configurations are labeled for brevity.}
    \label{configurations_ev_curves}
\end{figure}

The three lowest energy configurations are shown in \textbf{Figure~\ref{configurations_ev_curves}}. The ferromagnetic (FM) ground state is shown in \textbf{Figure~\ref{configurations_ev_curves}a}, where the magnetic moments on all \ch{Fe} atoms are spin-up. \textbf{Figure~\ref{configurations_ev_curves}b} and \textbf{Figure~\ref{configurations_ev_curves}c} depict the next two lowest energy configurations, labeled as SF28 and SF22, respectively. Here, SF denotes spin flipping, where the number of spin-up magnetic moments differs from that of spin-down moments, resulting in a net magnetic moment.   

The energy-volume curves representing $E_{k,0}(V)$ for the 25 magnetic configurations are shown in \textbf{Figure~\ref{configurations_ev_curves}d}, where only the three lowest energy configurations are labeled for brevity. Notice that all non-ground-state configurations have smaller equilibrium volumes than the ground state, giving rise to the negative thermal expansion shown in a later plot \cite{Liu2022ZentropyExpansion}.

The vibrational and electronic Helmholtz energy contributions, $F_{k,vib}(V,T)$ and $F_{k,el}(V,T)$, were added to $E_{k,0}(V)$ over the temperature range 0--1000 K to obtain $F_k(V,T)$ in Eq.~\eqref{eq:Fk_components}. The entropy $S_k(V, T)$ and specific heat capacity $C_{V, k}$ were obtained in a similar fashion by adding their vibrational and electronic contributions. As the resulting $F_k(V, T)$ data were fitted to the third-order or four-parameter Birch–Murnaghan equation of state, its first and second derivatives were readily available. These constitute the main inputs to \texttt{pyzentropy}. 

\subsection{Results and Discussion}
\begin{figure}[!htbp]
    \centering
    \includegraphics[width=1\textwidth, trim=0.1cm 0.3cm 0.1cm 0.1cm, clip]{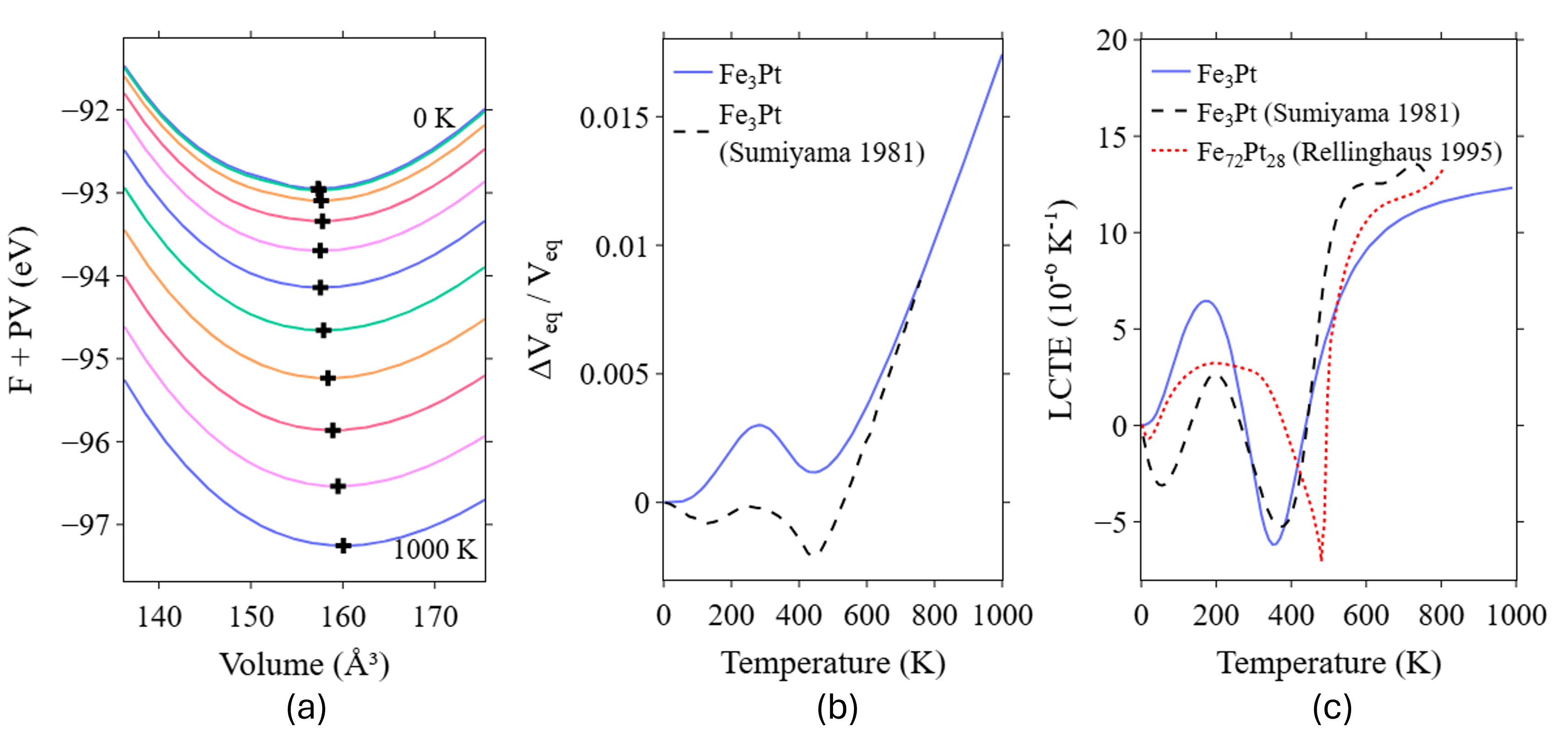}
    \caption{All results are at $P = 0$ GPa. (a) $F + PV$ vs. $V$ for fixed-$T$ curves from 0–1000 K in 100 K increments; $\fplus$ denotes the minima. (b) $\Delta V_{eq}/V_{eq}$ vs.\ $T$, compared with experimental results from Sumiyama et al. \cite{Sumiyama1981MagneticBoundary}. The reference $V_{eq}$ for this work corresponds to the 0 K value, while the experimental value is at 4.2 K. (c) LCTE vs.\ $T$, compared with experimental results from Sumiyama et al. \cite{Sumiyama1981MagneticBoundary} and Rellinghaus et al. \cite{Rellinghaus1995ThermodynamicInvar} for a slightly different composition of \ch{Fe72Pt28}. LCTE for Sumiyama et al. \cite{Sumiyama1981MagneticBoundary} was obtained from its $V_{eq}$ data by taking the derivative and fitting to a spline. The anomalous dip occurs in the vicinity of the second-order phase transition.}
    \label{Helmholtz_Veq_LCTE}
\end{figure}

\textbf{Figure~\ref{Helmholtz_Veq_LCTE}a} shows $F + PV$ for fixed-$T$ curves from 0–1000 K at $P = 0$ GPa. The minima on each curve are highlighted, and their $x$-values correspond to $V_{eq}$ shown in \textbf{Figure~\ref{Helmholtz_Veq_LCTE}b}, where the normalized values $\Delta V_{eq}/V_{eq}$ are plotted versus $T$. The simulation results reproduce the trend of the experimental data from Sumiyama et al.\ \cite{Sumiyama1981MagneticBoundary}, with the overall shape being similar, although the fractional change remains lower than the simulations up to approximately 800 K. 

\textbf{Figure~\ref{Helmholtz_Veq_LCTE}c} shows the corresponding LCTE (linear CTE, or CTE/3) results, in good agreement with the experimental data of Sumiyama et al. \cite{Sumiyama1981MagneticBoundary} and Rellinghaus et al. \cite{Rellinghaus1995ThermodynamicInvar} for a slightly different composition, \ch{Fe72Pt28}. The differences among the experimental results and the simulation can be attributed primarily to variations in composition, as an increase in Pt content is known to raise the Curie temperature $T_C$, which corresponds to a second-order phase transition at $P = 0$ GPa. This shifts the anomalous dip, which occurs near $T_C$, reported by Rellinghaus et al. \cite{Rellinghaus1995ThermodynamicInvar}, to a higher temperature ($\approx 480$ K) relative to the results of Sumiyama et al. ($\approx 370$ K) and the present simulation ($\approx 350$ K) \cite{Sumiyama1981MagneticBoundary}. The reported Curie temperatures are 505 K, 443 K, and 408 K corresponding to Rellinghaus et al. \cite{Rellinghaus1995ThermodynamicInvar}, Sumiyama et al. \cite{Sumiyama1981MagneticBoundary}, and this work, respectively. The order parameter is comparable across the studies (0.9 \cite{Rellinghaus1995ThermodynamicInvar}, 0.92 \cite{Sumiyama1981MagneticBoundary}, and 1), suggesting it does not significantly influence the observed behavior.

\begin{figure}[!htbp]
    \centering
    \includegraphics[width=1\textwidth, trim=0.1cm 0.3cm 0.1cm 0.1cm, clip]{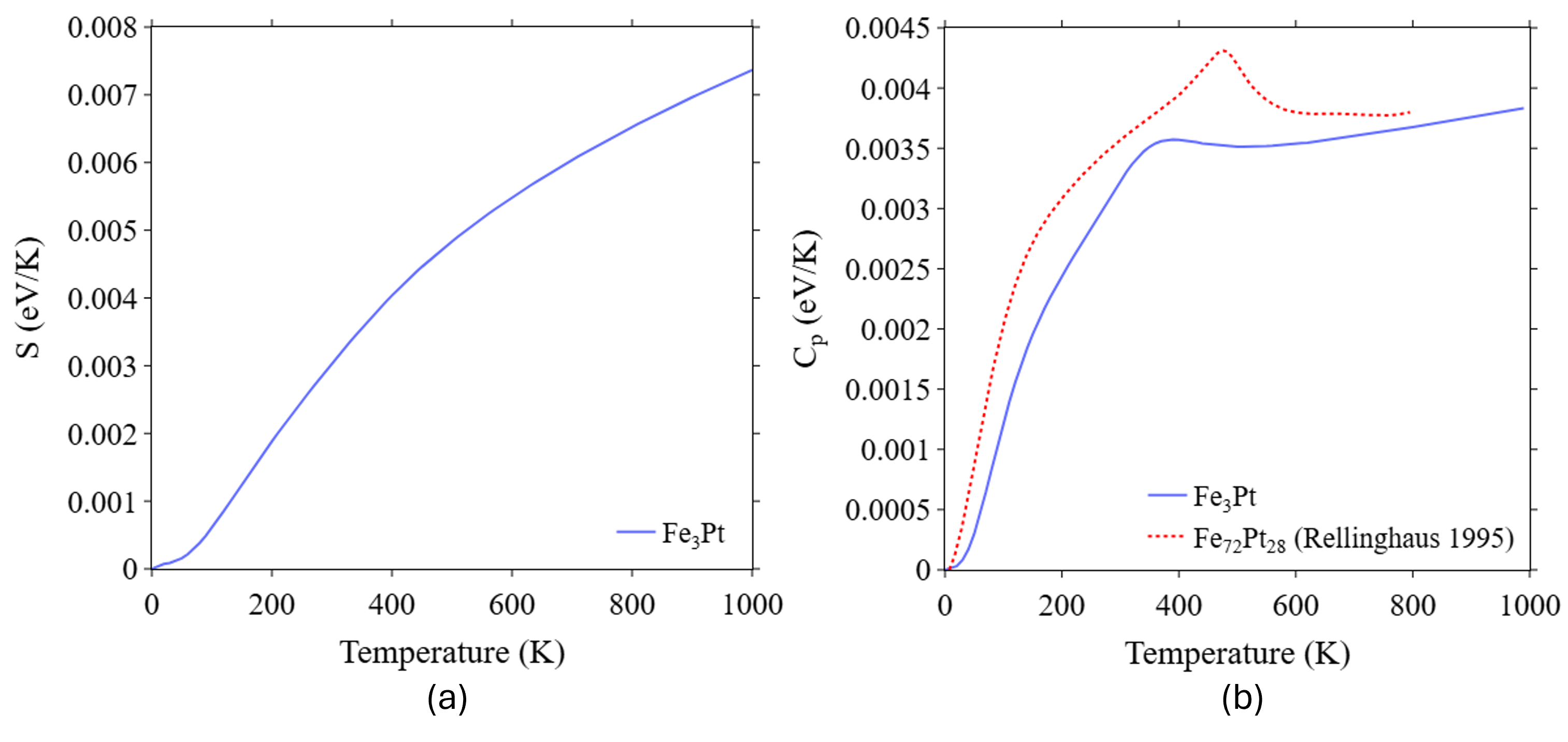}
    \caption{All results are at $P = 0$ GPa. (a) $S$ vs. $T$. (b) $C_P$ vs. $T$, compared with experimental results from Rellinghaus et al. \cite{Rellinghaus1995ThermodynamicInvar} for a slightly different composition, \ch{Fe72Pt28}. The anomalous peak occurs in the vicinity of the second-order phase transition.}
    \label{S_Cp}
\end{figure}

\textbf{Figure~\ref{S_Cp}a} presents the entropy $S$, while \textbf{Figure~\ref{S_Cp}b} shows the corresponding heat capacity $C_P$. Similar to the experimental results of Rellinghaus et al. \cite{Rellinghaus1995ThermodynamicInvar}, the simulations show a peak in the heat capacity in the vicinity of $T_C$, albeit not as pronounced. This result may be due to finite-size effects, as the limited number of sampled magnetic configurations broadens the peak and reduces its height compared to the experiment. In addition, the higher Pt content in the sample reported by Rellinghaus et al. \cite{Rellinghaus1995ThermodynamicInvar} shifts the peak to a higher temperature ($\approx 475$ K) compared to the simulation result ($\approx 390$ K). An additional peak which corresponds to the order–disorder transition has been reported at 1022 K, where the order parameter drops to zero \cite{Rellinghaus1995ThermodynamicInvar}, but predicting this transition is beyond the scope of the present work.

\begin{figure}[!htbp]
    \centering
    \includegraphics[scale=0.4, trim=0.1cm 0.3cm 0.1cm 3.1cm, clip]{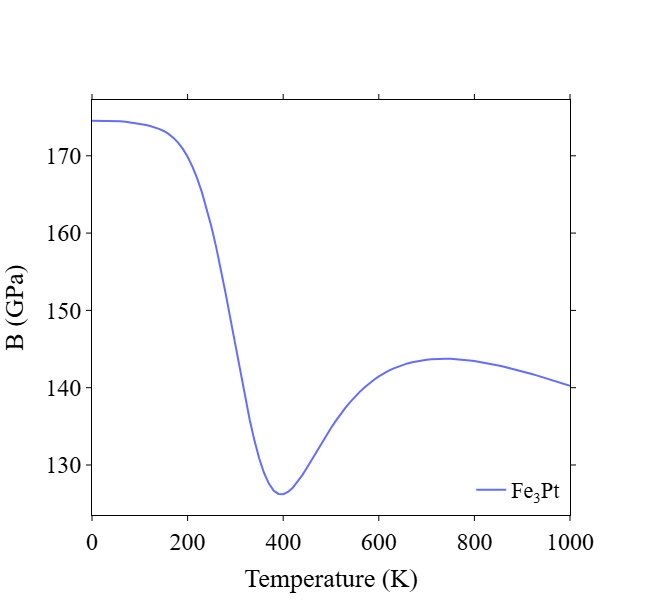}
    \caption{Plot of $B$ vs. $T$ at $P = 0$ GPa. The negative anomalous peak occurs in the vicinity of the second-order phase transition.}
    \label{bulk_modulus}
\end{figure}

The anomalous behavior is also evident in the bulk modulus plot shown in \textbf{Figure~\ref{bulk_modulus}}, where a dip occurs near the second-order phase transition ($\approx 400$ K). Similar temperature-dependent softening has been observed in other materials, such as \ch{CeBe13} \cite{Lenz1984TheCeBe13}. Thus, by using the recursive formula to calculate the total entropy considering different magnetic configurations, we were able to successfully predict the Invar behavior, as well as the anomalous temperature dependence of the LCTE, $C_P$, and $B$. It is important to note that considering only a single configuration does not reproduce these results.

\begin{figure}[!htbp]
    \centering
    \includegraphics[width=1\textwidth, trim=0.1cm 0.3cm 0.1cm 0.1cm, clip]{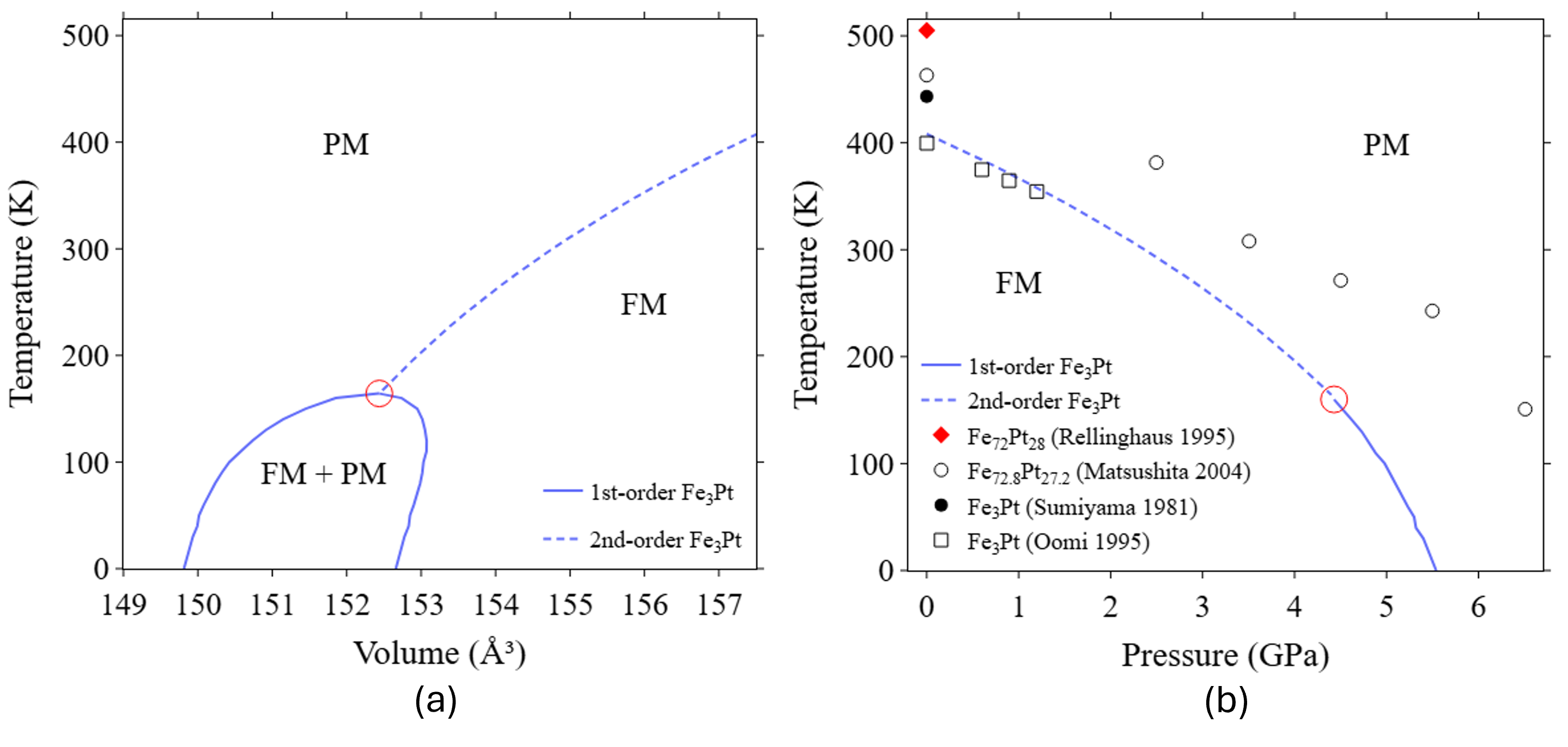}
    \caption{In both the (a) $T$–$V$ and (b) $T$–$P$ phase diagrams, the solid line represents the first-order transition (miscibility gap of the two-phase mixture), and the dotted line represents the second-order transition. In (b), the experimental points show the Curie temperatures $T_C$ for Fe$_{72}$Pt$_{28}$ (Rellinghaus et al. \cite{Rellinghaus1995ThermodynamicInvar}), Fe$_{72.8}$Pt$_{27.2}$ (Matsushita et al. \cite{Matsushita2004Pressure-inducedAlloy}), and Fe$_3$Pt (Sumiyama et al. \cite{Sumiyama1981MagneticBoundary}; Oomi and Araki \cite{Oomi1995EffectOrder}). The corresponding order parameters are 0.9, 0.85, 0.92, and 0.7, respectively. The red open circle denotes the tri-critical point.}
    \label{phase_diagrams}
\end{figure}

The results above correspond to a single pressure, $P = 0$ GPa. By performing calculations at multiple pressures to locate the second-order phase transition and constructing the common tangent of $F(V, T)$ to determine the first-order transition, the $T$–$V$ and $T$–$P$ phase diagrams can be constructed, as shown in \textbf{Figure~\ref{phase_diagrams}a} and \textbf{Figure~\ref{phase_diagrams}b}, respectively. The first-order transition corresponds to the miscibility gap of the two-phase FM + PM mixture. Here, PM denotes the paramagnetic phase, whose onset occurs when the ground-state spin probability $p_{GS} \le 0.5$, indicating sufficient spin disorder for the paramagnetic phase to appear; future work is needed to confirm that the total magnetization is zero. The tri-critical point occurs at $T = 160$ K, $V = 164.29$ \AA$^3$, and $P = 4.43$ GPa, beyond which the second-order phase transition follows.

The variation in the experimental results shown in \textbf{Figure~\ref{phase_diagrams}b} can be attributed to differences in composition and order parameter. At $P = 0$ GPa, $T_C$ is highest for Rellinghaus et al. \cite{Rellinghaus1995ThermodynamicInvar}, which has the highest Pt content (28\%) and an order parameter of 0.9, yielding $T_C = 505$ K. Matsushita et al. \cite{Matsushita2004Pressure-inducedAlloy}, with slightly lower Pt content (27.2\%) and an order parameter of 0.85, exhibits $T_C = 463$ K, 42 K lower. For Fe$_3$Pt, Sumiyama et al. \cite{Sumiyama1981MagneticBoundary} and Oomi and Araki \cite{Oomi1995EffectOrder} report $T_C = 443$ K and 400 K, respectively, with the lower value for Oomi and Araki \cite{Oomi1995EffectOrder} corresponding to a smaller order parameter of 0.7. 

Comparing our simulations for Fe$_3$Pt to those of Sumiyama et al. \cite{Sumiyama1981MagneticBoundary}, $T_C$ is in good agreement, being underpredicted by 35 K. The above comparisons explicitly refer to $P = 0$ GPa, as not all experimental studies report $T_C$ at elevated pressures. In general, our results reproduce the experimental trend when accounting for compositional variations and differences in order parameters.

\begin{figure}[!htbp]
    \centering
    \includegraphics[width=1\textwidth, trim=0.1cm 0.3cm 0.1cm 0.1cm, clip]{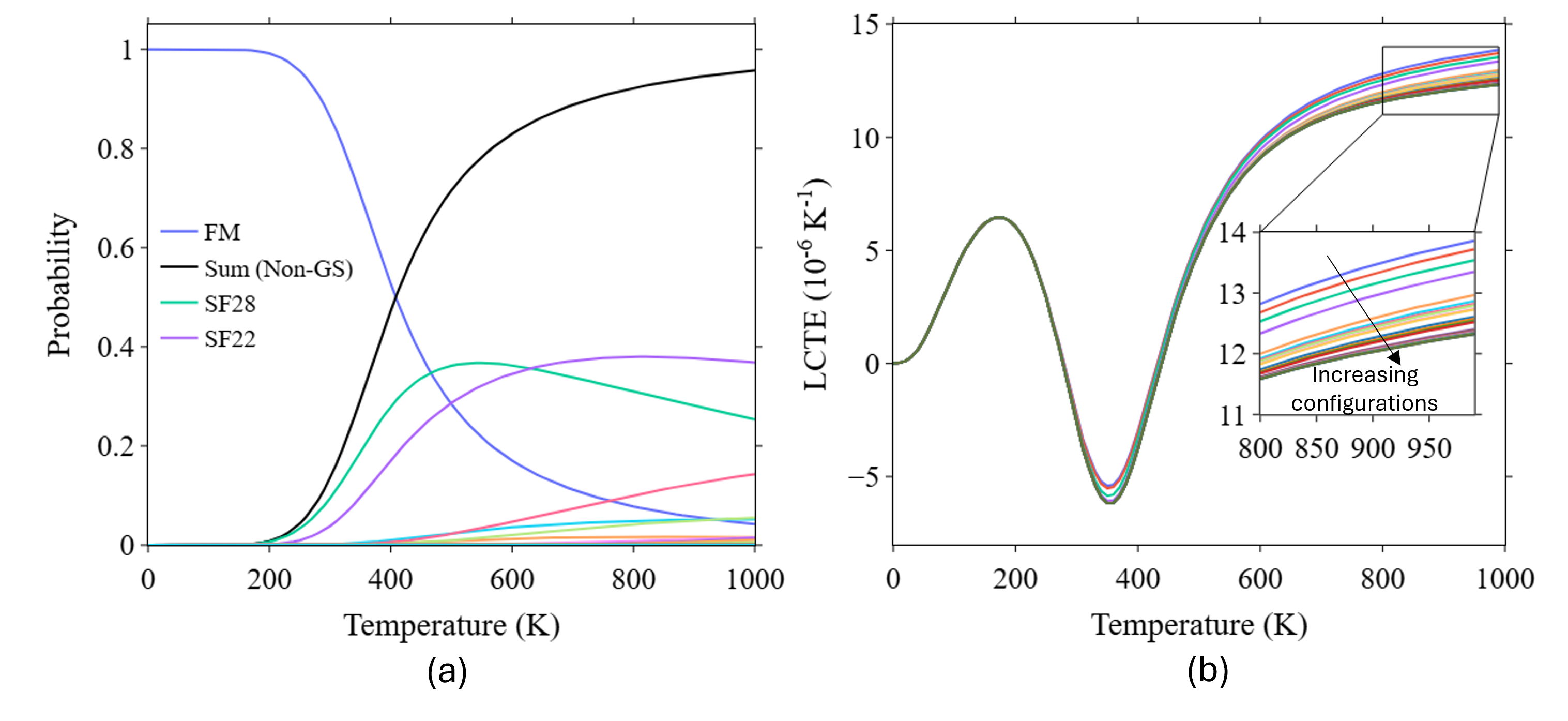}
    \caption{All results are at $P = 0$ GPa. (a) $p_k$ vs. $T$. The 3 dominant configurations are FM, SF28, and SF22, while the remaining configurations have $p_k < 0.2$ up to 1000 K. The solid black line represents the sum of $p_k$ over all non-ground-state configurations. (b) LCTE vs. $T$ for 3 to 25 configurations in unit increments. The topmost curve corresponds to the 3 dominant configurations, while the bottommost curve corresponds to all 25 configurations.}
    \label{probability_convergence}
\end{figure}

\FloatBarrier
All previous results were generated using 25 configurations. \textbf{Figure~\ref{probability_convergence}a} shows the configuration probabilities $p_k$ as a function of $T$ at $P = 0$ GPa. It can be seen that 3 configurations — the ground-state FM, SF28, and SF22 — dominate the probability distribution, while the remaining configurations have $p_k < 0.2$ up to 1000 K. Using the LCTE plot in \textbf{Figure~\ref{probability_convergence}b} as an example, where results are shown for 3 to 25 configurations in one configuration increments, it is evident that the 3 dominant configurations (topmost curve) are sufficient, particularly at lower temperatures ($< 600$ K). Only at higher temperatures do the remaining configurations contribute more significantly. Even so, including all 25 configurations (bottommost curve) shifts the LCTE by only $\sim1.5 \times 10^{-6}\ \mathrm{K}^{-1}$ at 990 K compared to the 3 dominant configurations. 

This study demonstrated the zentropy approach for a relatively small 12-atom \ch{Fe3Pt} supercell, which has 512 possible configurations, of which 37 are symmetrically unique. The limited number of unique configurations makes this study tractable using DFT. However, doubling the supercell to 24 atoms would dramatically increase the number of possible configurations to $2^{18} = 262{,}144$. This combinatorial explosion would limit zentropy studies using DFT alone. As noted previously, it is sufficient to sample only configurations with high probability. Thus, it is the interest of future work to employ methods such as Monte Carlo simulations based on cluster expansion or machine-learning interatomic potentials (MLIPs) to efficiently sample only configurations with sufficiently high probability, enabling zentropy studies of larger supercells and more complex alloys in practical case studies.

\section{Conclusion}
In conclusion, we introduced \texttt{pyzentropy}, an open-source Python package for implementing recursive entropy in first-principles thermodynamics. Using a 12-atom \ch{Fe3Pt} supercell with multiple magnetic configurations as a case study, we successfully captured its Invar behavior, as well as the anomalous temperature dependence of the LCTE, $C_P$, and $B$. We further constructed $T$–$V$ and $T$–$P$ phase diagrams. The results showed good agreement with experimental observations, while acknowledging current limitations due to finite-size effects. Finally, we emphasized the importance of including dominant configurations with sufficiently high probabilities, as lower-probability configurations contribute less significantly to the resulting material properties.

Future work will focus on applying \texttt{pyzentropy} to a broader range of systems, including other magnetic materials, larger supercells, and more complex alloys. To address the combinatorial explosion of possible configurations, we suggest combining the zentropy approach with complementary methods, such as Monte Carlo simulations based on cluster expansion and MLIPs, to efficiently identify the most relevant configurations. Future versions of the package will aim to enhance user-friendliness and incorporate new functionalities.

\section*{Acknowledgments} 
This work was supported by the Department of Energy (DOE) with Award Nos. DE-AR0001435 and DE-NE0009288, and by the National Science Foundation with Award Nos. CMMI-2226976 and CMMI-2050069. 

Computational resources were provided by the Bridges-2 supercomputer at the Pittsburgh Supercomputing Center and the Stampede3 supercomputer at the Texas Advanced Computing Center through project DMR140063: Fundamental Properties of Crystalline Materials by First Principles from the Advanced Cyberinfrastructure Coordination Ecosystem: Services \& Support (ACCESS) program~\cite{10.1145/3569951.3597559}, which is supported by U.S. National Science Foundation grants \#2138259, \#2138286, \#2138307, \#2137603, and \#2138296. The authors also recognize the Penn State Institute for Computational and Data Sciences (ICDS)  (RRID:SCR\_025154) for providing access to computational research infrastructure (RRID:SCR\_026424), and this research was partially supported by the Rising Researcher award ICDS\_RRYY\_027521 from Penn State’s ICDS (RRID:SCR\_025154).

\section*{CRediT author statement}
\textbf{Nigel Lee En Hew:} Conceptualization, Methodology, Software, Validation, Formal analysis, Investigation, Data Curation, Writing - Original Draft, Writing - Review \& Editing, Visualization. 

\textbf{Luke Allen Myers:} Conceptualization, Validation, Writing - Review \& Editing. 

\textbf{Shun-Li Shang:} Conceptualization, Methodology, Software, Writing - Review \& Editing, Supervision, Funding acquisition. 

\textbf{Zi-Kui Liu:} Conceptualization, Supervision, Funding acquisition.

\section*{Data Availability}
The \texttt{pyzentropy} package is publicly available on GitHub \cite{Pyzentropy}, where the \ch{Fe3Pt} example is included, albeit limited to 3 configurations. A Jupyter Notebook demonstrating all 25 configurations, as well as the generation of data and figures used in this manuscript, is available through the hew-publications GitHub organization \cite{NigelLeeEnHew2026Hew-publications}.

\bibliographystyle{elsarticle-num}  
\bibliography{references}           

@inproceedings{10.1145/3569951.3597559,
    title = {{ACCESS: Advancing Innovation: NSF’s Advanced Cyberinfrastructure Coordination Ecosystem: Services {\&} Support}},
    year = {2023},
    booktitle = {Practice and Experience in Advanced Research Computing 2023: Computing for the Common Good},
    author = {Boerner, Timothy J and Deems, Stephen and Furlani, Thomas R and Knuth, Shelley L and Towns, John},
    pages = {173–176},
    series = {PEARC '23},
    publisher = {Association for Computing Machinery},
    url = {https://doi.org/10.1145/3569951.3597559},
    address = {New York, NY, USA},
    isbn = {9781450399852},
    doi = {10.1145/3569951.3597559}
}

@article{Kresse1996EfficiencySet,
    title = {{Efficiency of ab-initio total energy calculations for metals and semiconductors using a plane-wave basis set}},
    year = {1996},
    journal = {Computational Materials Science},
    author = {Kresse, G and Furthm{\"{u}}ller, J},
    number = {1},
    month = {7},
    pages = {15--50},
    volume = {6},
    url = {https://www.sciencedirect.com/science/article/pii/0927025696000080},
    doi = {https://doi.org/10.1016/0927-0256(96)00008-0},
    issn = {0927-0256}
}

@article{Kresse1996EfficientSet,
    title = {{Efficient iterative schemes for ab initio total-energy calculations using a plane-wave basis set}},
    year = {1996},
    journal = {Physical Review B},
    author = {Kresse, G and Furthm{\"{u}}, J and Furthm{\"{u}}ller, J},
    number = {16},
    month = {10},
    pages = {11169--11186},
    volume = {54},
    publisher = {American Physical Society},
    url = {https://link.aps.org/doi/10.1103/PhysRevB.54.11169},
    doi = {10.1103/PhysRevB.54.11169},
    issn = {0163-1829}
}

@article{Togo2023First-principlesPhono3py,
    title = {{First-principles Phonon Calculations with Phonopy and Phono3py}},
    year = {2023},
    journal = {Journal of the Physical Society of Japan},
    author = {Togo, Atsushi},
    number = {1},
    volume = {92},
    doi = {10.7566/JPSJ.92.012001},
    issn = {0031-9015}
}

@article{Kresse1999FromMethod,
    title = {{From ultrasoft pseudopotentials to the projector augmented-wave method}},
    year = {1999},
    journal = {Physical Review B},
    author = {Kresse, G and Joubert, D},
    number = {3},
    pages = {1758--1775},
    volume = {59},
    publisher = {American Physical Society},
    url = {https://link.aps.org/doi/10.1103/PhysRevB.59.1758},
    doi = {10.1103/PhysRevB.59.1758}
}

@article{Perdew1996GeneralizedSimple,
    title = {{Generalized Gradient Approximation Made Simple}},
    year = {1996},
    journal = {Physical Review Letters},
    author = {Perdew, John P and Burke, Kieron and Ernzerhof, Matthias},
    number = {18},
    pages = {3865--3868},
    volume = {77},
    doi = {10.1103/PhysRevLett.77.3865},
    issn = {0031-9007}
}

@article{Sumiyama1979CharacteristicAlloys,
    title = {{Characteristic magnetovolume effects in Invar type Fe-Pt alloys}},
    year = {1979},
    journal = {Journal of Physics F: Metal Physics},
    author = {Sumiyama, K and Shiga, M and Morioka, M and Nakamura, Y},
    number = {8},
    pages = {1665},
    volume = {9},
    url = {https://dx.doi.org/10.1088/0305-4608/9/8/017},
    doi = {10.1088/0305-4608/9/8/017},
    issn = {0305-4608}
}

@article{Hew2025DensityApproximation,
    title = {{Density Functional Theory ToolKit (DFTTK) to automate first-principles thermodynamics via the quasiharmonic approximation}},
    year = {2025},
    journal = {Computational Materials Science},
    author = {Hew, Nigel Lee En and Myers, Luke Allen and van de Walle, Axel and Shang, Shun-Li and Liu, Zi-Kui},
    number = {114072},
    month = {8},
    pages = {114072},
    volume = {258},
    url = {https://www.sciencedirect.com/science/article/pii/S092702562500415X https://linkinghub.elsevier.com/retrieve/pii/S092702562500415X},
    doi = {10.1016/j.commatsci.2025.114072},
    issn = {09270256}
}

@misc{DFTTK,
    title = {{DFTTK}},
    url = {https://github.com/PhasesResearchLab/dfttk}
}

@article{Oomi1995EffectOrder,
    title = {{Effect of pressure on the Curie temperature of Fe3Pt alloys having different degree of order}},
    year = {1995},
    journal = {Journal of Magnetism and Magnetic Materials},
    author = {Oomi, Gendo and Araki, Hiroyuki},
    month = {2},
    pages = {83--84},
    volume = {140-144},
    doi = {10.1016/0304-8853(94)01148-6},
    issn = {03048853}
}

@article{Shang2010First-principlesNi3Al,
    title = {{First-principles thermodynamics from phonon and Debye model: Application to Ni and Ni3Al}},
    year = {2010},
    journal = {Computational Materials Science},
    author = {Shang, Shun-Li and Wang, Yi and Kim, DongEung and Liu, Zi-Kui},
    number = {4},
    month = {2},
    pages = {1040--1048},
    volume = {47},
    url = {https://www.sciencedirect.com/science/article/pii/S0927025609004558},
    doi = {https://doi.org/10.1016/j.commatsci.2009.12.006},
    issn = {0927-0256}
}

@incollection{Bengtsson2017GeometryDistributions,
    title = {{Geometry of probability distributions}},
    year = {2017},
    booktitle = {Geometry of Quantum States},
    author = {Bengtsson, I. and {\.{Z}}yczkowski, K.},
    editor = {Bengtsson, Ingemar and {\.{Z}}yczkowski, Karol},
    edition = {2},
    month = {8},
    pages = {29--62},
    publisher = {Cambridge University Press},
    url = {https://www.cambridge.org/core/product/C29D42D6C8D2A5499AA05B145820D94D https://www.cambridge.org/core/product/identifier/CBO9781139207010A015/type/book_part},
    address = {Cambridge},
    isbn = {9781107656147},
    doi = {10.1017/9781139207010.003}
}

@misc{NigelLeeEnHew2026Hew-publications,
    title = {{hew-publications}},
    year = {2026},
    author = {{Nigel Lee En Hew}},
    url = {https://github.com/hew-publications}
}

@article{Togo2023ImplementationPhono3py,
    title = {{Implementation strategies in phonopy and phono3py}},
    year = {2023},
    journal = {Journal of Physics: Condensed Matter},
    author = {Togo, Atsushi and Chaput, Laurent and Tadano, Terumasa and Tanaka, Isao},
    number = {35},
    month = {9},
    pages = {353001},
    volume = {35},
    doi = {10.1088/1361-648X/acd831},
    issn = {0953-8984}
}

@article{Perdew2021InterpretationsTheories,
    title = {{Interpretations of ground-state symmetry breaking and strong correlation in wavefunction and density functional theories}},
    year = {2021},
    journal = {Proceedings of the National Academy of Sciences},
    author = {Perdew, John P. and Ruzsinszky, Adrienn and Sun, Jianwei and Nepal, Niraj K. and Kaplan, Aaron D.},
    number = {4},
    month = {1},
    volume = {118},
    doi = {10.1073/pnas.2017850118},
    issn = {0027-8424}
}

@book{Dove1993IntroductionDynamics,
    title = {{Introduction to Lattice Dynamics}},
    year = {1993},
    author = {Dove, Martin T.},
    month = {10},
    publisher = {Cambridge University Press},
    isbn = {9780521392938},
    doi = {10.1017/CBO9780511619885}
}

@article{Sumiyama1981MagneticBoundary,
    title = {{Magnetic and Magnetovolume Properties of the Cu3Au Type Ordered Fe-Pt Alloys around the {$\gamma$}- {$\alpha$} Phase Boundary}},
    year = {1981},
    journal = {Journal of the Physical Society of Japan},
    author = {Sumiyama, Kenji and Emoto, Yoshiaki and Shiga, Masayuki and Nakamura, Yoji},
    number = {10},
    month = {10},
    pages = {3296--3302},
    volume = {50},
    publisher = {The Physical Society of Japan},
    url = {https://doi.org/10.1143/JPSJ.50.3296},
    doi = {10.1143/JPSJ.50.3296},
    issn = {0031-9015}
}

@article{Blochl1994ProjectorMethod,
    title = {{Projector augmented-wave method}},
    year = {1994},
    journal = {Physical Review B},
    author = {Bl{\"{o}}chl, P E},
    number = {24},
    pages = {17953--17979},
    volume = {50},
    publisher = {American Physical Society},
    url = {https://link.aps.org/doi/10.1103/PhysRevB.50.17953},
    doi = {10.1103/PhysRevB.50.17953}
}

@article{Shang2023QuantifyingFe3Pt,
    title = {{Quantifying the degree of disorder and associated phenomena in materials through zentropy: Illustrated with Invar Fe3Pt}},
    year = {2023},
    journal = {Scripta Materialia},
    author = {Shang, Shun-Li and Wang, Yi and Liu, Zi-Kui},
    month = {3},
    pages = {115164},
    volume = {225},
    publisher = {Pergamon},
    url = {https://www.sciencedirect.com/science/article/pii/S1359646222006595},
    doi = {https://doi.org/10.1016/j.scriptamat.2022.115164},
    issn = {1359-6462}
}

@book{beale2011statistical,
    title = {{Statistical Mechanics}},
    year = {2011},
    author = {Beale, P D},
    publisher = {Academic Press},
    url = {https://books.google.com/books?id=KdbJJAXQ-RsC},
    isbn = {9780123821898}
}

@article{Liu2022ZentropyExpansion,
    title = {{Zentropy Theory for Positive and Negative Thermal Expansion}},
    year = {2022},
    journal = {JOURNAL OF PHASE EQUILIBRIA AND DIFFUSION},
    author = {Liu, Zi-Kui and Wang, Yi and Shang, Shun-Li},
    number = {6},
    month = {12},
    pages = {598--605},
    volume = {43},
    doi = {10.1007/s11669-022-00942-z},
    issn = {1547-7037}
}

@article{Matsushita2004Pressure-inducedAlloy,
    title = {{Pressure-induced change of the magnetic state in ordered Fe–Pt Invar alloy}},
    year = {2004},
    journal = {Journal of Magnetism and Magnetic Materials},
    author = {Matsushita, M. and Endo, S. and Miura, K. and Ono, F.},
    number = {3},
    month = {3},
    pages = {393--397},
    volume = {269},
    doi = {10.1016/S0304-8853(03)00648-6},
    issn = {03048853}
}

@misc{Pyzentropy,
    title = {{pyzentropy}},
    url = {https://github.com/PhasesResearchLab/pyzentropy}
}

@misc{NigelLeeEnHew2026PyzentropyDocumentation,
    title = {{pyzentropy Software Documentation}},
    year = {2026},
    author = {{Nigel Lee En Hew}},
    url = {https://pyzentropy.readthedocs.io/en/main/}
}

@article{Myers2025RecursiveApproach,
    title = {{Recursive entropy in thermodynamics: establishing the statistical-physics basis of the zentropy approach}},
    year = {2025},
    journal = {arXiv:2511.04950 [cond-mat.stat-mech]},
    author = {Myers, Luke Allen and Lee, Nigel and Hew, En and Shang, Shun-Li and Liu, Zi-Kui},
    month = {11},
    url = {https://arxiv.org/pdf/2511.04950},
    arxivId = {2511.04950}
}

@misc{TomPreston-WernerSemantic2.0.0,
    title = {{Semantic Versioning 2.0.0}},
    author = {{Tom Preston-Werner}},
    url = {https://semver.org/}
}

@book{Grosso2014SolidPhysics,
    title = {{Solid State Physics}},
    year = {2014},
    author = {Grosso, Giuseppe and Parravicini, Giuseppe Pastori},
    publisher = {Elsevier},
    isbn = {9780123850300},
    doi = {10.1016/C2010-0-66724-1}
}

@article{Lenz1984TheCeBe13,
    title = {{The anomalous bulk modulus of CeBe13}},
    year = {1984},
    journal = {Solid State Communications},
    author = {Lenz, D. and Schmidt, H. and Ewert, S. and Boksch, W. and Pott, R. and Wohlleben, D.},
    number = {8},
    month = {11},
    pages = {759--763},
    volume = {52},
    doi = {10.1016/0038-1098(84)90405-8},
    issn = {00381098}
}

@article{Rellinghaus1995ThermodynamicInvar,
    title = {{Thermodynamic analysis of Fe72Pt28 Invar}},
    year = {1995},
    journal = {Physical Review B},
    author = {Rellinghaus, Bernd and K{\"{a}}stner, Jochen and Schneider, Thomas and Wassermann, Eberhard F and Mohn, Peter},
    number = {5},
    month = {2},
    pages = {2983--2993},
    volume = {51},
    publisher = {American Physical Society},
    url = {https://link.aps.org/doi/10.1103/PhysRevB.51.2983},
    doi = {10.1103/PhysRevB.51.2983}
}

@article{Wang2010ThermodynamicPrototype,
    title = {{Thermodynamic fluctuations in magnetic states: Fe3Pt as a prototype}},
    year = {2010},
    journal = {Philosophical Magazine Letters},
    author = {Wang, Yi and Shang, Shun-Li and Zhang, H. and Chen, Long-Qing and Liu, Zi-Kui},
    number = {12},
    month = {12},
    pages = {851--859},
    volume = {90},
    url = {https://www.tandfonline.com/doi/full/10.1080/09500839.2010.508446},
    doi = {10.1080/09500839.2010.508446},
    issn = {09500839}
}

\newpage
\appendix
\renewcommand\thefigure{\thesection.\arabic{figure}}
\setcounter{figure}{0}
\renewcommand{\thetable}{A\arabic{table}}
\setcounter{table}{0}

\end{document}